\documentclass[aps,pra,showpacs,superscriptaddress,twocolumn]{revtex4}

\usepackage{amsmath}
\usepackage{amssymb}
\usepackage{graphicx}
\usepackage{dcolumn}
\usepackage{bm} 
\usepackage{epstopdf}

\newcommand{\ket}[1]{|{#1}\rangle}
\newcommand{\bra}[1]{\langle{#1}|}
\newcommand{\ketbras}[3]{\ket{#1}_{#3}\hspace*{-0.mm}\bra{#2}}
\newcommand{\Tr}{\mathrm{\text{Tr}}}
\newcommand{\bracket}[2]{\langle#1|#2\rangle}

\begin{document}
\title{Quantum parameter estimation affected by unitary disturbance}

\author{A. De Pasquale}
\affiliation{NEST, Scuola Normale Superiore and Istituto Nanoscienze-CNR, I-56126 Pisa, Italy} 

\author{D. Rossini}
\affiliation{NEST, Scuola Normale Superiore and Istituto Nanoscienze-CNR, I-56126 Pisa, Italy} 

\author{P. Facchi}
\affiliation{Dipartimento di Fisica and MECENAS, Universit\`a di Bari, and INFN Sezione di Bari, I-70126 Bari, Italy} 

\author{V. Giovannetti}
\affiliation{NEST, Scuola Normale Superiore and Istituto Nanoscienze-CNR, I-56126 Pisa, Italy} 


\begin{abstract}
  We provide a general framework for handling the effects of a unitary disturbance 
  on the estimation of the amplitude $\lambda$ associated to a unitary dynamics.
  By computing an analytical and general expression for the quantum Fisher information, 
  we prove that the optimal estimation precision for $\lambda$ cannot be outperformed 
  through the addition of such a unitary disturbance. 
  However, if the dynamics of the system is already affected by an external field, 
  increasing its strength does not necessary imply a loss in the optimal estimation precision. 
\end{abstract}



\maketitle
\section{introduction}
The quantum Cram\'{e}r-Rao bound provides the proper theoretical framework for analyzing energy/time-like 
uncertainty relations~\cite{BRAU} by setting limits on the precision attainable when estimating 
the parameters governing the dynamics of a physical system.
Its application has profound consequences in quantum metrology~\cite{quantumetrology}, 
where it helps in identifying which resources (e.g., entanglement and squeezing) are useful 
to reach higher accuracy levels, and which are the proper procedures one needs to adopt 
to fully exploit them. For instance, in the absence of external noise, 
the quantum Cram\'{e}r-Rao bound predicts~\cite{METRO} that in the process of
estimating a relative phase, the use of entanglement between sequences of $N$ independent 
probing systems allows one to gain a $\sqrt{N}$ improvement in precision (Heisenberg limit) 
over those procedures which, under the same experimental settings, adopt instead separable probes
(standard quantum limit, or shot-noise limit in optical interferometry). 
More subtle is to establish the optimal performances in presence of external disturbances. 
Many discouraging results attest the fragility of entanglement which, 
in a noisy environment, limits any precision improvement at most to a constant 
factor independent of $N$~\cite{SQL}, or to a super-classical precision scaling, 
$N^{-5/6}$, achieved when the perturbation involves a preferential direction perpendicular 
to the unitary evolution governed by the parameter to be estimated~\cite{acin}. 
Yet an exhaustive answer would require a systematic method for taking 
into account the presence of disturbance in the system. The main obstacle is represented 
by the fact that the very fundamental tool needed to evaluate the quantum Cram\'{e}r-Rao inequality, 
{\it i.e.}, the quantum Fisher information (QFI)~\cite{fisher25}, apart from very simple cases, 
usually happens to be computationally cumbersome, especially for high dimensional systems. 
Recently, Escher \textit{et al.}~\cite{escher} proposed a strategy to circumvent 
this difficulty by introducing an upper bound to the QFI, which relies on the choice 
of a Kraus representation of the noisy evolution based on physical considerations. 
In this way, it was possible to propose a realistic example of optical lossy interferometry 
where the Heisenberg limit can be attained by properly tuning 
the number of input resources according to the noise level~\cite{escher}. 
A generalization of the latter analysis for the single parameter estimation to the case of lossy 
optical waveform reconstruction has been recently proposed~\cite{tsangOpt}. 
Notwithstanding these bright techniques, a general prescription for computing 
the QFI of a generic dynamical process is still missing.

In this paper we consider the case of closed quantum systems and study the effects 
of a unitary disturbance on the estimation of a dynamical parameter $\lambda$. 
Specifically, we add a term to the generator of the dynamical evolution of the system 
to model the action of an external force that opposes the formation of the parametric trajectories 
governed by $\lambda$, and determine a compact analytical expression for the associated QFI. 
Starting from this result, we report several important facts. 
First of all, while for non optimal choices of model settings it is possible that the disturbance 
will improve the accuracy of the estimation procedure, we prove a \textit{No-Go theorem} 
which formalizes the rather intuitive fact that the {\it best} performances are always 
reached when no disturbance is present in the system. Most importantly we also notice that enhancing 
the level of a Hamiltonian disturbance which is 
already affecting the system does \emph{not} necessarily yield a worse {\it optimal} estimation strategy 
and can reveal itself helpful in determining the value of $\lambda$. 
This is a rather counterintuitive finding which can be interpreted as the emergence 
of {\it dithering}~\cite{DITHER} in the estimation process. 

This paper is organized as follows. In Sec.~\ref{sec:QET}, after briefly reviewing 
the typical approach followed for the reconstruction of the global phase $\lambda$ 
of a generic unitary evolution, we explicitly address the case in which the dynamics 
is affected by the presence of a unitary disturbance (Sec.~\ref{sec:QET_unidisturb}). 
This technique is then reframed in the more general context of multiparametric estimation 
(Sec.~\ref{sec:QET_multip}). 
In Sec.~\ref{sec:nogoth}, we prove a \textit{No-go theorem} comparing the optimal 
performances achievable with and without the unitary disturbance. 
In Sec.~\ref{sec:nogoth_heisenberg} we address the question of whether the additional term 
in the Hamiltonian is sufficient to induce a departure from the Heisenberg limit, 
and show that this is not the case. 
Finally, in Sec.~\ref{sec:dithering}, we specialize to the case of a single qubit, 
and gather evidence that if the system is already affected by such unitary disturbance, 
the latter can be increased in order to achieve better estimation performances. 
Sec.~\ref{sec:conclutions} is devoted to final remarks.

\section{Theoretical framework}\label{sec:QET}

A standard problem in quantum estimation theory is recovering 
a real parameter $\lambda$ encoded in a set of states $\varrho_\lambda$ of the system. 
The ultimate precision limit for such task is given by the quantum 
Cram\'er-Rao bound~\cite{quantumetrology} on the Root Mean Square Error (RMSE) $\Delta \lambda$ of a generic estimation 
strategy (the latter is defined as
$\Delta \lambda = \sqrt{ \mathbb{E}[(\lambda^{(\rm est)} - \lambda)^2]}$
where $\lambda^{(\rm est)}$ is the random variable which represents the estimation of $\lambda$ 
extrapolated from the performed measurements and $\mathbb{E}[x]$ indicates
the expectation value of the random variable $x$). Accordingly, we have 
\begin{equation} \label{QFIBOUND}
  \Delta \lambda \geq {1}/{\sqrt{\nu {\cal Q}}},
\end{equation} 
where ${\cal Q}$ is the QFI obtained by optimizing the Fisher information~\cite{CRAMER} 
over all the possible positive-operator valued measurements performed on the system encoding the parameter, 
and $\nu$ is the number of times the measurement is repeated (the threshold being reachable 
at least in the asymptotic limit of large $\nu$ --- see however Ref.~\cite{HAYASHI} 
for achievability at finite $\nu$). 
The QFI is a function of the parameter $\lambda$ which can be expressed 
in terms of the ``instantaneous" velocity variation of the system, quantified by 
the Bures distance $\mathcal{D}_{B}$~\cite{bures}, 
\begin{equation}
  {\cal Q} = 4 \lim_{\delta \lambda \to 0} \frac{\mathcal{D}_B^2 (\varrho_{\lambda},\varrho_{\lambda+\delta \lambda})}{\delta \lambda^2}= 
  8 \lim_{\delta \lambda \to 0} \frac{1-\mathcal{F}(\varrho_{\lambda},\varrho_{\lambda+\delta \lambda})}{\delta \lambda^2}\;,
  \label{eq:QFI}
\end{equation} 
where $\mathcal{F}(\varrho,\varrho') = \Tr \big[ \sqrt{\sqrt{\varrho} \, \varrho' \sqrt{\varrho}} \big]$ 
is the fidelity between the states $\varrho$ and $\varrho'$~\cite{fidelity}.

A well studied case is the one in which the parameter $\lambda$ is encoded into the state 
of a quantum system through a unitary transformation of the form
\begin{equation}
  \varrho_\lambda = U_\lambda \varrho_0 U_\lambda^\dagger, \qquad U_{\lambda} =\exp( -i \lambda H_I),
  \label{eq:undyn}
\end{equation}
where $\varrho_0$ is the input state of the system, assumed to be controllable, 
and $H_I$ is the generator of the parametric orbit, assumed to be assigned and independent 
from $\lambda$ (as an example, consider the case of a massive particle that undergoes 
an abrupt translation induced by an external force whose intensity we wish to estimate 
by monitoring the particle).
Under these conditions the QFI is also independent from $\lambda$, and is given by~\cite{paris,caves} 
\begin{equation}
  {\cal Q}[\varrho_0] = 4\sum_{j<j'} \frac{(\rho_j - \rho_{j'})^2}{\rho_j + \rho_{j'}}|\bra{j} H_I \ket{j'}|^2\,,
  \label{eq:QFI_nodisturb}
\end{equation}
where $\rho_j$ and $|j\rangle$ are respectively the eigenvalues and the eigenvectors of $\varrho_0$. 
From the strong concavity of ${\cal F}$ it follows that the maximum of Eq.~(\ref{eq:QFI_nodisturb}) 
is obtained for pure states $\varrho_0 = \ket{\phi_0}\bra{\phi_0}$.
In that case Eq.~(\ref{eq:QFI_nodisturb}) becomes
\begin{equation}
  {\cal Q} = 4 \bra{\phi_0} \Delta^2 H_I \ket{\phi_0} 
  = 4 \left(\bra{\phi_0} H_I^2 \ket{\phi_0}-\bra{\phi_0} H_I\ket{\phi_0}^2\right)\,.
\label{eq:Qpure}
\end{equation} 
Thereby the state $\varrho_0$ maximizing the value of ${\cal Q}$ 
[{\it i.e.}, minimizing the RMSE threshold~(\ref{QFIBOUND})] 
can be identified by observing that the maximum variance of the Hermitian operator $H_I$ 
is proportional to the square of its spectral width:
\begin{equation}
  \langle\Delta^2 H_I \rangle_{\max}=\frac{(h_{\max}-h_{\min})^2}{4}\,,
\end{equation}
where
$h_{\max/\min}$ is the maximum/minimum eigenvalue of $H_I$.
Accordingly we have
\begin{equation}{\cal Q}_{\max} :=
  \max_{\varrho_0} {\cal Q}[\varrho_0]= (h_{\max}-h_{\min})^2\;,
\label{eq:Qpureoptimal}
\end{equation}
the maximum being achieved by using as optimal input 
$\varrho_0^{\rm(opt)} =|\psi_0^{\rm(opt)}\rangle\langle \psi_0^{\rm(opt)}|$
the equally weighted superposition 
of the eigenvectors belonging to $h_{\max/\min}$, {\it i.e.}
\begin{equation}
  |\psi_0^{\rm(opt)}\rangle = \frac{1}{\sqrt{2}} (\ket{h_{\max}}+\ket{h_{\min}}) \; .
\end{equation}

\subsection{Phase estimation with unitary disturbance}\label{sec:QET_unidisturb}

Let us now consider the case where the above estimation process is disturbed by the presence of an 
additional contribution to the generator of the dynamics. 
Specifically we replace $U_\lambda$ of Eq.~(\ref{eq:undyn}) with the transformation 
\begin{equation}
  U_{\lambda,\eta} = \exp[-i H(\lambda,\eta)], \qquad
  H(\lambda,\eta) = \lambda H_I + \eta H_0,
\end{equation}
where $H_0$ is an Hermitian operator interfering with the parametric driving exerted by $H_I$, 
and where the real quantity $\eta$ gauges the strength of the associated perturbation
(in the example discussed previously, $H_0$ can be identified 
with a trapping potential that opposes the translation of the massive particle).
In order to compute the QFI for $\lambda$, for any fixed $\eta$, we apply Uhlmann's 
theorem on the fidelity~\cite{uhlmann}:
\begin{equation}
 \mathcal{F}(\varrho_\lambda, \varrho_{\lambda+\delta \lambda}) 
= \max_{\ket{\varrho_{\lambda}},\ket{\varrho_{\lambda+\delta \lambda}}} |\bracket{\varrho_\lambda}{\varrho_{\lambda+\delta \lambda}}|\ ,
\end{equation}
the maximization being performed over all possible purifications $\ket{\varrho_{\lambda}}$ 
and $\ket{\varrho_{\lambda+\delta \lambda}}$ of $\varrho_{\lambda}$ and $\varrho_{\lambda+\delta \lambda}$, 
respectively, through an ancillary system.
By using the freedom in the purifications we write 
\begin{equation}
  \mathcal{F} = \max_{V} \Big| \bra{\varrho_0} \, \overleftarrow{\exp}
  \left[ -i \, \delta \lambda \, \bar{H}_I(\lambda,\eta) \right] \otimes V \ket{\varrho_0} \Big| ,
\end{equation}
where $V$ belongs to the set of unitary transformations on the ancilla, while 
$\ket{\varrho_0} = \sum_{j} \sqrt{\rho_j} \ket{j}\otimes \ket{j}$ is a fixed purification 
of the initial state $\varrho_0$ 
(hereafter, by writing $A \otimes B$ we mean that $A$ acts on the system and $B$ on the ancilla). 
The average Hamiltonian 
 \begin{equation}
  \bar{H}_I (\lambda,\eta) = \int_{0}^1 dt \, e^ {i H(\lambda,\eta) t} H_I e^ {-i H (\lambda,\eta) t}
  \label{eq:barHI}
\end{equation}
emerges from the interaction picture representation of the evolution
\begin{equation}
U_{\lambda + \delta \lambda, \eta} = 
U_{\lambda, \eta} \overleftarrow{\exp} [-i \, \delta \lambda \, \bar{H}_I(\lambda, \eta)]\ ,
\end{equation}
with $\overleftarrow{\exp}[\ldots]$ denoting the time-ordered exponential
(a similar approach was employed in Ref.~\cite{tsang}).
Since we are interested in the limit of small $\delta \lambda$, without loss of generality 
we set $V=\exp(i \, \delta \lambda \, \Omega)$, with $\Omega$ a Hermitian operator 
on the ancillary system. It results that, up to corrections of order $O(\delta \lambda^4)$, 
the fidelity reads
\begin{equation}
  \mathcal{F} \simeq 1-\frac{\delta \lambda^2}{2} \min_\Omega \Big[ \bra{\varrho_0} ( \Delta \bar{H}_I \otimes \mathbb{I} - \mathbb{I} \otimes \Delta \Omega )^2 \ket{\varrho_0} \Big]\,,
  \label{eq:fidelity}
\end{equation}
where $\Delta \bar{H}_I = \bar{H}_I-\bar{h}$ and $\Delta \Omega = \Omega-\omega$, 
with $\bar{h} = \Tr[\varrho_0 \bar{H}_I]$ and $\omega = \Tr[\varrho_0\Omega]$. 
Using the spectral decomposition of $\varrho_0$ introduced above, 
the QFI in Eq.~(\ref{eq:QFI}) can be written as
\begin{eqnarray}\label{eq:QFImin}
  {\cal Q}_{\lambda; \eta} [\varrho_0]&=&4 \min_\Omega \Tr\Big[ \bar{H}_I^2\varrho_0 
    + \Omega^2 \varrho_0 -2\sqrt{\varrho_0} \bar{H}_I^\top \sqrt{\varrho_0}\Omega  \nonumber\\
    && - {\bar{h}}^2 \varrho_0 - \sum_{i,j} \rho_i \rho_j  \Omega \ketbras{i}{j}{}\Omega \ketbras{j}{i}{}
    +2  \bar{h} \varrho_0 \Omega\Big]\,, \nonumber\\
\end{eqnarray}
where $\top$ denotes transposition.
By differentiating the trace with respect to $\Omega$ we determine the minimization condition for it:
\begin{equation}\label{eq:eqForOmega}
  \varrho_0 \left(\Omega - \omega \right) +  \left(\Omega - \omega\right)\varrho_0 
  = 2 \sqrt{\varrho_0} \left(\bar{H}_I^\top - \bar{h} \right ) \sqrt{\varrho_0}\,.
\end{equation}
Its solution displays a  translational invariance with respect to $\omega$ according to
\begin{equation}
  \Omega - \omega   = \Omega'- \omega' , \quad \mbox{with}\quad \Omega' =\Omega+g \mathbb{I},~g\in \mathbb{C}\,.
\end{equation} 
Without loss of generality we can therefore fix $\omega=\bar{h}$ and write the solution 
of Eq.~(\ref{eq:eqForOmega}) in a basis for the ancilla isomorphic to the eigenbasis of $\varrho_0$ as
\begin{equation}
  \Omega_{jj'} = 2 [\bar{H}_I]_{j'j} \frac{\sqrt{ \rho_{j} \rho_{j'} }} {\rho_{j} +\rho_{j'}}\ .
\end{equation}
Finally, by substituting this solution into~(\ref{eq:QFImin}), 
we obtain the QFI for $\lambda$ in the presence of an arbitrary disturbance $\eta H_0$: 
\begin{equation}
  {\cal Q}_{\lambda; \eta} [\varrho_0] = 4\sum_{j<j'}\frac{(\rho_j-\rho_{j'})^2}{\rho_j+\rho_{j'}}|\bra{j}\bar{H}_I\ket{j'}|^2\,.
  \label{eq:QFImixedlambda}
\end{equation}
Notice that for $\eta=0$, since $\bar{H}_I$ reduces to $H_I$, this expression gives back 
Eq.~(\ref{eq:QFI_nodisturb}),  {\it i.e.}, ${\cal Q}_{\lambda; 0} [\varrho_0] = {\cal Q}[\varrho_0]$.
Furthermore in complete analogy to the latter case, if the initial state of the system is pure, 
Eq.~(\ref{eq:QFImixedlambda}) yields
\begin{equation}
  {\cal Q}_{\lambda;\eta} [\ket{\phi_0}] = 4\bra{\phi_0}\Delta^2 \bar{H}_I \ket{\phi_0}.
  \label{eq:puredist}
\end{equation} 
At variance with Eqs.~(\ref{eq:QFI_nodisturb}) and~(\ref{eq:Qpure}), for $\eta\neq 0$, 
Eqs.~(\ref{eq:QFImixedlambda}) and~(\ref{eq:puredist}) can exhibit an explicit dependence 
on $\lambda$ via Eq.~(\ref{eq:barHI}) (an example is provided below). 
In particular, this implies that the optimal states $\varrho_0^{\rm(opt)}$ yielding 
the maximum of the QFI (and of course the QFI maximum itself), can now depend on
the value of the parameter one wishes to estimate. 
Specifically, indicating with $\bar{h}_{\max/\min}(\lambda,\eta)$ the maximum/minimum eigenvalue 
of the average Hamiltonian $\bar{H}_I(\lambda,\eta)$ and with $\ket{\bar{h}_{\max/\min}}$ 
its corresponding eigenvector, we now get 
\begin{equation} 
  {\cal Q}_{\lambda;\eta}^{(\max)}:= \max_{\varrho_0} {\cal Q}_{\lambda;\eta}[\varrho_0]
  = [\bar{h}_{\max}(\lambda,\eta)-\bar{h}_{\min}(\lambda,\eta)]^2,
  \label{eq:QpureoptimalNEW}
\end{equation} 
with the optimal state $\varrho_0^{\rm(opt)} =|\psi_0^{\rm(opt)}\rangle\langle \psi_0^{\rm(opt)}|$ 
being the superposition~\cite{NOTA1}
\begin{equation}
  |\psi_0^{\rm(opt)}\rangle=\frac{1}{\sqrt{2}}(\ket{\bar{h}_{\max}(\lambda,\eta)}+\ket{\bar{h}_{\min}(\lambda,\eta)}) \ .
\end{equation}

\subsection{Multiparametric estimation}\label{sec:QET_multip}

Equations~(\ref{eq:QFImixedlambda}) and~(\ref{eq:QpureoptimalNEW}) represent the central 
finding of our paper, and pave the way to a number of observations on the role played 
by a unitary disturbance in the estimation procedure.
Before detailing  them, we notice that the analysis presented so far can be naturally framed 
in the more general context of multiparametric estimation, where the family of states 
$\varrho_{\vec{\lambda}}$ now depends on a set of parameters $\vec{\lambda}=(\lambda_1,\dots,\lambda_M)$, 
with $M\geq 2$.
In this context the Cram\'er-Rao theorem is generalized to a bound 
\begin{equation}
  \bm{\mathrm{Cov}}[\vec{\lambda}] \geq \bm{\mathcal Q}^{-1}/\nu
\end{equation}
on the covariance matrix, 
\begin{equation}
  \big[ \bm{\mathrm{Cov}}[\vec{\lambda}] \big]_{j,k} = \mathbb{E}[\lambda^{(\rm est)}_j \lambda^{(\rm est)}_k]- \lambda_j \lambda_k \ , 
\end{equation}
with $\bm{\mathcal Q}$ being the QFI matrix of the problem. 
While referring the reader to the formal expression of $\bm{\mathcal Q}$, we remind that 
its diagonal elements coincide with the QFI of the corresponding parameter $\lambda_j$, 
at fixed values of the others. 
The off-diagonal terms can be evaluated in a similar way by observing that for
any other set of parameters $\vec{\mu}=\vec{\mu}(\vec{\lambda})$, 
which is an invertible function of $\vec{{\lambda}}$, the associated QFI matrix 
can be computed as ${\bm{\tilde {\mathcal Q}}} = \bm{J}{\bm{\mathcal Q}}{\bm{J}^\top}$, 
where $\bm{J}$ is the Jacobian matrix with elements $[\bm{J}]_{jk}=\partial \lambda_k/\partial \mu_j$. 

Let us consider, for example, the case of two parameters $\vec{\lambda}=(\lambda,\eta)$. 
The diagonal elements become, respectively, 
\begin{equation}
[\bm{\mathcal Q}]_{\lambda\lambda } = {\cal Q}_{\lambda;\eta}[\varrho_0],
\qquad 
[\bm{\mathcal Q}]_{\eta\eta} = {\mathcal Q}_{\eta;\lambda}[\varrho_0], 
\end{equation}
given by Eq.~(\ref{eq:QFImixedlambda}) and its analog obtained by substituting 
$\bar{H}_I$ with $\bar{H}_0$ [whose definition is exactly specular to that in Eq.~(\ref{eq:barHI})]. 
By defining
\begin{equation}
  \mu_{1} =\frac{\lambda + \eta}{\sqrt{2}}\;, \qquad  \quad \mu_{2} =\frac{\lambda - \eta}{\sqrt{2}}\ , 
\end{equation}
one can also compute $[\bm{\mathcal Q}]_{\lambda \eta}$ as 
\begin{equation}
  [\bm{\mathcal Q}]_{\lambda \eta} = [\bm{\mathcal Q}]_{\eta\lambda} 
  = [\bm{\tilde{\mathcal Q}}]_{\mu_1\mu_1} - \big( [\bm{\mathcal Q}]_{\lambda \lambda} + [\bm{\mathcal Q}]_{\eta\eta} \big) / 2.
\end{equation} 
From the previous analysis of the QFI at $M=1$ for a system affected by a unitary disturbance, 
it follows that $[\bm{\tilde{\mathcal Q}}]_{\mu_1\mu_1}$ can be smoothly determined 
by rewriting the global Hamiltonian of the system as 
\begin{equation}
  H=\mu_1 \frac{(H_I+ H_0)}{\sqrt{2}}+\mu_2 \frac{(H_I- H_0)}{\sqrt{2}} \ ,
\end{equation}
and by using Eq.~(\ref{eq:QFImixedlambda}) upon substituting $\bar{H}_I$ with $(\bar{H}_I+\bar{H}_0)/\sqrt{2}$. 
It follows that the off-diagonal terms of the QFI matrix are 
\begin{equation} 
  [\bm{\mathcal Q}]_{\lambda \eta} 
  = 4\sum_{j<j'} \frac{(\rho_j-\rho_{j'})^2}{\rho_j+\rho_{j'}} 
  \mathrm{Re} \big[\bra{j}\bar{H}_I \ket{j'} \bra{j'} \bar{H}_0{} \ket{j} \big] \,.
\end{equation}
This technique can be naturally extended to the case of an arbitrary number of parameters.

\section{No-Go theorem}\label{sec:nogoth}
A question which spontaneously arises 
from the similarity between the expressions for the QFI with and without a unitary disturbance, 
{\it i.e.}, Eqs.~(\ref{eq:QFI_nodisturb}) and~(\ref{eq:QFImixedlambda}), concerns the possibility 
to compare the performances of an estimation procedure in the two cases. 
First of all, it is evident that for non optimal choices of the input state $\varrho_0$, 
it is indeed possible that a non-zero value of $\eta$ could help the estimation process 
(for an explicit example, take $\varrho_0$ to be an eigenvector of $H_I$~\cite{NOTAexample}).
However, in terms of {\it optimal} estimation thresholds, the following No-Go theorem can be derived:

\textit{No-go theorem}. It is not possible to outperform the optimal estimation strategy for the amplitude $\lambda$
of the unitary dynamics~(\ref{eq:undyn}) through the addition of any linear contribution 
to its generator, namely,
\begin{equation}
{\cal Q}^{(\max)}_{\lambda;\eta}
  \leq {\cal Q}^{(\max)}_{\lambda;0}= {\cal Q}_{\max}\,. 
  \label{eq:noGo}
\end{equation}
In order to prove this inequality we observe that Eqs.~(\ref{eq:Qpureoptimal}) 
and~(\ref{eq:QpureoptimalNEW}) allow us to equivalently rewrite it in terms of a contraction 
of the spectral width of the Hamiltonian $\bar{H}_I$ with respect to ${H}_I$, {\it i.e.}, 
\begin{equation}
 \bar{h}_{\max}(\lambda,\eta)-\bar{h}_{\min}(\lambda,\eta) \leq h_{\max}-h_{\min}\ . 
 \end{equation}
The latter can then be proved by observing that the operator $\bar{H}_I$ is obtained 
from $H_I$ via a weighted convex sum of random unitaries. 
Therefore, according to Uhlmann's majorization theorem~\cite{nielsen}, $\bar{H}_I$ is majorized by $H_I$. 
This in particular implies
\begin{equation}
 \bar{h}_{\max}(\lambda,\eta)
\leq \bar{h}_{\max}(\lambda,0) \quad \mbox{and} \quad \bar{h}_{\min}(\lambda,\eta) \geq \bar{h}_{\min}(\lambda, 0)
\end{equation}
from which the contraction of the spectral width, and hence Eq.~(\ref{eq:noGo}), is derived.

\subsection*{Heisenberg limit}\label{sec:nogoth_heisenberg}

Once established that ${\cal Q}^{(\max)}_{\lambda;\eta}$ is always smaller than ${\cal Q}^{(\max)}_{\lambda;0}$, 
one might ask whether or not this implies a departure from the Heisenberg limit of the optimal accuracy. 
We remind the reader that the latter is associated with the case where (say) $N$ independent probes
are prepared in entangled states before being acted upon by the generator of the dynamics. 
Formally this can be accounted for by replacing $H_I$ of Eqs.~(\ref{eq:undyn}) to~(\ref{eq:Qpure}) 
with the operator
\begin{equation}
H_I^{(N)} = \sum_{j=1}^N H_I^{(j)} 
\end{equation}
with $H_I^{(j)}$ being the local generator 
acting the $j$-th probe (see Ref.~\cite{METRO} for details). As a result, ${\cal Q}_{\max}$ 
of Eq.~(\ref{eq:Qpureoptimal}) becomes
\begin{equation}
{\cal Q}^{(N)}_{\max}= N^2 (h_{\max}-h_{\min})^2
\end{equation}
with $h_{\max/\min}$ being still the extremal eigenvalues of the local (single probe) Hamiltonian $H_I$ 
(the $N^2$ dependence certifying the arising of the Heisenberg limit in the RMSE accuracy). 
If the Hamiltonian disturbance $H_0$ is acting locally on the individual probes 
it is immediate to see that the same dependence upon $N$ remains also for $\eta \neq 0$. 
Indeed in this case $H(\lambda,\eta)$ is replaced by 
\begin{equation}
  H^{(N)}(\lambda,\eta) = \sum_{j=1}^N (\lambda H_I^{(j)} + \eta H_0^{(j)}) 
\end{equation}
which is still given 
by a sum of $N$ independent, local, contributions yielding 
\begin{equation}
  {\cal Q}_{\lambda;\eta}^{(N,\max)} = N^2 [\bar{h}_{\max}(\lambda,\eta)-\bar{h}_{\min}(\lambda,\eta)]^2 \, ,
\end{equation} 
where bars refer to eigenvalues of the single probe Hamiltonian $\bar{H}_I$.
The situation becomes more complex when $H_0$ is non-local and $H^{(N)}(\lambda,\eta)$ 
acquires coupling terms between the $N$ probes. 
As the overall evolution is still unitary, one is tempted to conjecture that the same $N^2$ 
scaling for ${\cal Q}_{\lambda;\eta}^{(N,\max)}$ should survive in typical situations.
A rigorous proof of this fact is left to a future investigation.

\section{Optimal accuracy improvement via disturbance}\label{sec:dithering}
Inequality~(\ref{eq:noGo}) compares the best achievable 
performance with and without the addition of a linear disturbance $\eta H_0$ 
to the generator of a given unitary dynamics~(\ref{eq:undyn}). 
From this relation one could be tempted to conclude that the maximum QFI 
is a monotonic decreasing function of $\eta$; that is, the larger the disturbance is, 
the worse the estimation of $\lambda$. 
In general, however, this is not true: once the threshold
$\eta=0$ has been crossed, Eq.~(\ref{eq:noGo}) does not provide 
a recipe for comparing the response of the system to increasing or decreasing values 
of $\eta$~\cite{NOTAMAJ}. This opens the possibility of dithering effects.

We now provide an explicit example of such phenomenon in a qubit system. 
Let us adopt the Bloch sphere formalism and set 
\begin{equation}
  H_I = \bm{a} \cdot\bm{\sigma}, \qquad H_0=\bm{b}\cdot\bm{\sigma},
\end{equation}
where, without any loss of generality, $\bm{a}$ and $\bm{b}$ 
are unit (three-dimensional) real vectors, and $\bm{\sigma}$ is the vector of 
Pauli matrices. 
In this case the average Hamiltonian is given by $\bar{H}_I=\bm{m}\cdot \bm{\sigma}$, with
\begin{eqnarray}
  \bm{m} & = &
     [1+\mbox{sinc}(2\theta)] \bm a /2 - \eta (\bm b \wedge \bm a) \, \mbox{sinc}^2\theta \nonumber \\
     & & + \frac{1-\mbox{sinc}(2\theta)}{2\theta^2}\left[(\bm n \cdot \bm a) \bm n-\eta (\bm b \wedge \bm a)\wedge \bm n \right] \, ,
\end{eqnarray}
where $\wedge$ is the vector product, $\mbox{sinc}\, x = x^{-1} \sin x$, and
\begin{equation}
  \bm n = \lambda \bm a + \eta \bm b, \qquad \theta=|\bm n|.
\end{equation}
From Eq.~(\ref{eq:QpureoptimalNEW}) it immediately follows that  
 \begin{equation}
   {\cal Q}^{(\max)}_{\lambda;\eta} = 4|\bm{m}|^2\;. \label{NEWQ}
 \end{equation} 
 
\begin{figure}[!t]
  \includegraphics[height=0.62\columnwidth]{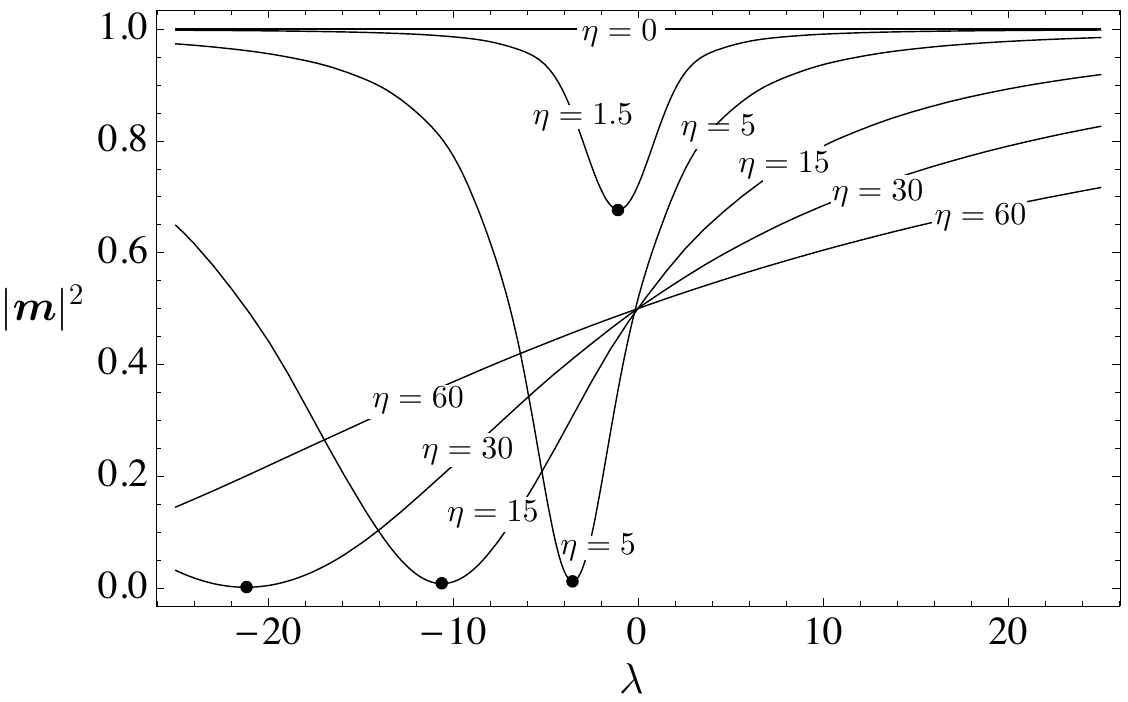}
  \caption{Plot of ${\cal Q}^{(\max)}_{\lambda;\eta}/4$ of Eq.~(\ref{NEWQ}) as a function of $\lambda$ 
    for different values of $\eta$, and for $\bm{a}\cdot\bm{b}=1/\sqrt{2}$. 
    The dots signal the minima at $\lambda_{\min}=-\eta \bm{a}\cdot \bm{b}$.}
  \label{fig:msquare}
\end{figure} 

In Fig.~\ref{fig:msquare} we plot Eq.~(\ref{NEWQ}) as a function of $\lambda$ 
for different values of $\eta \geq 0$, the case $\eta<0$ being symmetric with respect to $\lambda=0$. 
For $\eta=0$ (no unitary disturbance) we have $|\bm{m}|^2=1$ for all $\lambda$'s: 
as already observed, the optimal choice for the initial state of the system does not depend 
on the amplitude of the unitary dynamics. On the other hand, according to the No-Go theorem, 
for all $\eta\neq0$ we always have that $|\bm{m}|^2 < 1$.
This function shows a minimum at $\lambda_{\min}=-\eta \, \bm{a} \cdot \bm{b}$,
and asymptotically reaches $1$ for $\lambda\to\pm \infty$ (in this regime the effects 
of the unitary disturbance $\eta H_0$ can be considered negligible). 
The antilinear relation between $\lambda_{\min}$ and $\eta$, 
determines the following behavior of $|\bm{m}|^2$: 
for $\eta$ large enough, there exists an interval $I$ such that for $\lambda \in I$ 
\begin{equation}\label{INEQ}
  {\cal Q}^{(\max)}_{\lambda;\eta} < {\cal Q}^{(\max)}_{\lambda;\tilde{\eta}},\qquad \mathrm{for} \quad \tilde{\eta} > \eta
\end{equation} 
(see Fig.~\ref{fig:msquare}).
Inequality~(\ref{INEQ}) establishes that, for sufficiently large $\eta$, there exists 
a finite interval of $\lambda$'s whose values, by properly choosing the state of the input probe, 
can be estimated better than the values achievable with {\it any} possible choice 
of $\varrho_0$, when the unitary disturbance is smaller.

\section{Conclusions}\label{sec:conclutions}
The optimal estimation precision for the amplitude of a unitary dynamics 
cannot be enhanced by switching on an external field, or more generally by adding a linear term 
to the generator of the dynamical process. However, we have shown that 
if the system is already affected by such a unitary disturbance, enhancing its strength 
does not necessary imply a loss in the estimation precision of the other dynamical parameter(s). 
These results have been achieved by explicitly computing the quantum Fisher information 
for an arbitrary system in a generic mixed state, thus generalizing the already known expression 
for the case of a unitary dynamics~(\ref{eq:undyn}). 
Furthermore, reframed into the more general context of multiparametric estimation, 
this analysis enabled us to easily determine a compact analytical expression for all 
the elements of the quantum Fisher information matrix. 

\acknowledgments We thank R. Fazio and G. Florio for useful discussions. This work 
was supported by MIUR through FIRB-IDEAS Project No. RBID08B3FM, and by PRIN 2010LLKJBX.


\begin{thebibliography}{99}
\bibitem{NOTAexample}
  For such choice ${\cal Q}_{\lambda;0}[\varrho_0] =0$ implying that \emph{no} information 
  can be recovered on $\lambda$ when $\eta=0$ [the state of the probe being invariant 
  under the mapping~(\ref{eq:undyn})]. On the contrary, as in general for $\eta\neq 0$ 
  the operator $\bar{H}_I$ differs from $H_I$, the same result does not necessary hold 
  when the disturbance is on (indeed in this case we can have ${\cal Q}_{\lambda;0}[\varrho_0] >0$).
\bibitem{NOTAMAJ}
  Notice that, from the definition~(\ref{eq:barHI}) of $\bar{H}_I$, 
  we cannot assert that $\bar{H}_I(\lambda,\eta+\Delta \eta)$ majorizes $\bar{H}_I(\lambda,\eta)$ 
  for arbitrary $\eta$ and $\Delta \eta \neq 0$.
\end{thebibliography}
\end{document}